\documentclass[12pt]{article}
\usepackage{color}
\usepackage{amsmath}
\usepackage{cite}
\usepackage[pdftex]{graphicx}
\begin{document}    
\vspace*{1cm}

\renewcommand\thefootnote{\fnsymbol{footnote}}
\begin{center} 
  {\Large\bf Right-handed neutrino dark matter consistent with the generation of 
baryon number asymmetry}
\vspace*{1cm}

{\Large Daijiro Suematsu}\footnote[1] {professor emeritus, ~e-mail:
suematsu@hep.s.kanazawa-u.ac.jp}
\vspace*{0.5cm}\\

{\it Institute for Theoretical Physics, Kanazawa University, 
Kanazawa 920-1192, Japan}
\end{center}
\vspace*{1.5cm} 

\noindent
{\Large\bf Abstract}\\
A right-handed neutrino is a promising candidate for dark matter (DM) which has 
no interaction with nuclei. Since two right-handed neutrinos explain 
neutrino oscillation data and baryon number asymmetry through both the seesaw mechanism 
and leptogenesis based on the out-of-equilibrium decay of the lighter one, the lightest 
third right-handed neutrino could be free from them and stable to be DM. 
It suggests a possibility that the right-handed neutrinos could be common key 
particles for the explanation of three problems in the standard model (SM).
We study this possibility by analyzing the structure of a neutrino mass matrix in the
scotogenic model for neutrino mass. 

\newpage
\setcounter{footnote}{0}
\renewcommand\thefootnote{\alph{footnote}}

\section{Introduction}
Neutrino oscillation experiments prove the existence of nonzero neutrino mass \cite{noscil}. 
Observations of both the abundances of light elements and the cosmic microwave background 
suggest the existence of baryon number asymmetry in the Universe \cite{bbn,planck}. 
However, both the neutrino mass and the baryon number asymmetry cannot be explained
in the standard model (SM). It suggests that the SM is incomplete and has to be extended.
If the SM is extended by two right-handed neutrinos which are assumed to have Majorana mass,
these two problems are considered to be solved through the so called seesaw 
mechanism \cite{seesaw} and leptogenesis caused by the out-of-equilibrium decay of 
the right-handed neutrino \cite{leptg} as long as their Majorana mass is larger 
than $O(10^9)$ GeV \cite{lbound}.
   
On the other hand, the SM has no candidate of dark matter (DM) whose existence 
has been proved through various astrophysical and cosmological observations. 
Its direct detection experiments severely constrain its mass and elastic scattering cross section 
with a nucleus. As a result, a lot of DM candidates are excluded from their consistency 
with the required DM relic abundance \cite{dm}.
It may suggest a possibility that DM has no interaction with nuclei. 
In this sense, the right-handed neutrino, which is massive and electrically neutral, 
could be a good candidate of DM  since it has no interaction with nuclei.
A possibility such that a right-handed neutrino is DM has been studied from a various view points. 
One may find recent works in \cite{n1dm,rndm,nonext1,nonext2}.
If we combine this idea with the above solution for the neutrino mass and the baryon number asymmetry, 
the third right-handed neutrino seems to be a good DM candidate if its stability is 
guaranteed by some symmetry. 
This requirement for the symmetry may impose to introduce an additional doublet scalar $\eta$
since we have to guarantee the existence of their invariant Yukawa couplings between 
doublet leptons and right-handed neutrinos under the assumed symmetry. 
As such an example, we have the scotogenic model \cite{scot}, which can make 
the neutrino mass generation possible radiatively around TeV scales.
The model seems to be very interesting as a simple extension of the SM since
three big problems in the SM might be explained by the right-handed neutrinos simultaneously.  

However, if we consider such a possibility that the lightest right-handed neutrino $N_1$ is DM in the
scotogenic model, we have to answer the following hard problems simultaneously:
\begin{enumerate}
\item[(i)] How the required DM relic abundance is realized? 
If $N_1$ has only neutrino Yukawa couplings $h_{\alpha 1}\bar\ell_{L_\alpha}\eta N_1$, 
large values are required for $h_{\alpha 1}$ to reduce its abundance through the annihilation 
in the freezeout scenario. 
This makes it difficult to suppress lepton flavor violating processes (LFV) 
like $\mu\rightarrow e\gamma$ below the experimental bounds \cite{mks}.
\item[(ii)] How the leptogenesis can work successfully in a consistent way with the neutrino 
mass generation and the DM abundance? 
If the right-handed neutrinos are supposed to have TeV scale mass,
it seems to be difficult to generate sufficient lepton number asymmetry to explain 
the required baryon number asymmetry \cite{ks}.
\end{enumerate}
In previous studies aiming to solve these problems including the case where DM is the lightest
neutral component of $\eta$, 
the model is extended by some way,
for example,  new interaction is introduced to the right-handed neutrinos 
\cite{comm,u1,newi,zeromass}.  
In this paper, we propose a solution to these problems within the original scotogenic model 
without considering any extension of the model.\footnote{It is discussed that 
$N_1$ can be DM consistently with the LFV constraint by considering coannihilation 
among the right-handed neutrinos without any extension \cite{tribi}.}   

The remaining parts are organized as follows. In section 2, after we review the model briefly,
we discuss a neutrino mass matrix derived from it in detail, and flavor mixing and mass eigenvalues
are fixed. Using them, we study a solution consistent with both the relic DM abundance and the 
constraint from the lepton flavor violating processes.
In section 3, we discuss the consistency of the scenario with successful leptogenesis.
We summarize the paper in section 4.    

\section{Neutrino model with right-handed neutrino DM}
A model studied here is a simple extension of the SM with three right-handed neutrinos 
$N_k$ and an inert doublet scalar $\eta$, which is called the scotogenic model \cite{scot}.
It is assumed to be invariant under an imposed $Z_2$ symmetry. 
New fields $N_k$ and $\eta$ are assumed to have its odd parity and its even parity
 is assigned for all the SM contents.
The extended part of Lagrangian is characterized by additional terms
\begin{eqnarray}
-{\cal L}_Y&=&\sum_{\alpha,k}h_{\alpha k}\bar\ell_{L_\alpha}\eta N_k 
+\sum_k\frac{1}{2}M_ke^{i\gamma_k}\bar N_kN_k^c 
+{\rm h.c.},
\nonumber\\
V&=&\lambda_1(\phi^\dagger\phi)^2+\lambda_2(\eta^\dagger\eta)^2+
\lambda_3(\phi^\dagger\phi)(\eta^\dagger\eta)+\lambda_4(\phi^\dagger\eta)(\eta^\dagger\phi)
+\frac{\lambda_5}{2}\left[(\phi^\dagger\eta)^2 +{\rm h.c.}\right]  \nonumber\\
&+&m_\phi^2\phi^\dagger\phi
+m_\eta^2\eta^\dagger\eta,
\label{lag}
\end{eqnarray}
where $\phi$ is the ordinary Higgs doublet scalar, and $\lambda_{4,5}<0$ and $M_1<M_2<M_3$ 
are assumed. Neutrino Yukawa couplings $h_{\alpha k}$ are defined using a basis that
a mass matrix of right-handed neutrinos is diagonal.  
Coupling constants in the potential $V$ should satisfy
\begin{equation}
\lambda_{1,2}>0, \qquad \lambda_3+\lambda_4+\lambda_5>-\sqrt{\lambda_1\lambda_2}
\label{stab}
\end{equation} 
for the vacuum stability. 
If $\eta$ is assumed to have no vacuum expectation value (VEV), neutrino mass is not 
produced at tree level and the $Z_2$ symmetry is kept as an exact symmetry of the model.

After the electroweak symmetry breaking,  mass of the components of $\eta$ is 
fixed through the Higgs VEV as
\begin{equation}
M_{\eta^\pm}^2=m_\eta^2+\lambda_3\langle\phi\rangle^2,  \quad 
M_{\eta_R}^2=m_\eta^2+(\lambda_3+\lambda_4+ \lambda_5)\langle\phi\rangle^2, \quad
M_{\eta_I}^2=m_\eta^2+(\lambda_3+\lambda_4- \lambda_5)\langle\phi\rangle^2,
\label{etamass}
\end{equation} 
where $\eta^\pm$ stands for charged components. $\eta_R$ and $\eta_I$ are real and 
imaginary part of the neutral component, respectively. 
In this setting, a lightest right-handed neutrino $N_1$ or $\eta_R$ can be stable 
due to the $Z_2$ symmetry. 
We suppose that $N_1$ is DM and the baryon number asymmetry is generated 
by leptogenesis through the out-of-equilibrium decay of a heavier right-handed neutrino.  
In the following study, mass ordering of these extra particles is assumed to satisfy \cite{ds} 
\begin{equation}
M_1<M_{\eta_R}<M_2<M_3.
\label{massorder}
\end{equation}

\subsection{Neutrino mass and flavor mixing}
Although neutrino mass is forbidden at tree level in Lagrangian (\ref{lag}), small neutrino mass is induced 
through a one-loop diagram caused by the Higgs VEV $\langle\phi\rangle$ as
\begin{eqnarray}
&&({\cal M}_\nu)_{\alpha\beta}=\sum_k h_{\alpha k} h_{\beta k} \Lambda_ke^{i\gamma_k}
\equiv\sum_k\mu_{\alpha\beta}^k \Lambda_ke^{i\gamma_k} ,
\nonumber \\
&&\Lambda_k=\frac{\lambda_5\langle\phi\rangle^2}{16 \pi^2M_k}
  \left[\frac{M_k^2}{M_\eta^2-M_k^2}
    \left(1+\frac{M_k^2}{M_\eta^2-M_k^2}
    \ln\frac{M_k^2}{M_\eta^2}\right) \right],
\label{scot}
\end{eqnarray}
where $M_\eta^2=\frac{1}{2}(M_{\eta_R}^2+M_{\eta_I}^2)$.
 Since there is no other contribution to the neutrino mass in the model, it is instructive to 
note an interesting feature of this mass formula.

Its diagonalization matrix can be described by using the Yukawa couplings $h_{\alpha k}$ explicitly.
When a condition $\sum_\alpha(h_{\alpha j}h_{\alpha k})\propto\delta_{jk}$ is satisfied, which
corresponds to the condition that matrices $\mu^j$ and $\mu^k$ in eq.~(\ref{scot}) are commutable \cite{comm},
${\cal M}_\nu$ is diagonalized by a matrix $U$ as $U^T{\cal M}_\nu U={\rm diag}(m_1,m_2,m_3)$. 
This $U$ can be expressed as 
\begin{equation}
U=(U_1,U_2,U_3), \quad U_k^T\equiv (U_{ek}, U_{\mu k}, U_{\tau k}), \quad
U_{\alpha k}\equiv \frac{h_{\alpha k}}{h_k},
\label{yukawa}
\end{equation}
where $h_k$ is a real constant. 
An absolute value of the mass eigenvalue $m_k$ is fixed as
\begin{equation}
m_k=\left|\sum_\alpha h_{\alpha k}^2\right|\Lambda_k,
=p_k h_k^2\Lambda_k,
\label{nmasseig}
\end{equation}
where $p_k$ is defined as $p_k=\left|\sum_\alpha U_{\alpha k}^2\right|$.
Here, it should be noted that $\sum_\alpha U_{\alpha k}^2=1$ is not satisfied if the Yukawa coupling 
$h_{\alpha k}$ is complex. By using this $U$, the PMNS matrix \cite{pmns} can be written as 
$V_{\rm PMNS}=U_e^\dagger U P$ , 
where $U_e$ is a matrix which is used to diagonalize a charged lepton mass 
matrix and $P$ is a matrix of Majorana phases 
${\rm diag}(1,e^{i(\gamma_2-\gamma_1)},e^{i(\gamma_3-\gamma_1)})$.
Although $V_{\rm PMNS}$ is unitary in a case where $h_{\alpha k}$ 
is real, it is not unitary in a case where $h_{\alpha k}$ is complex since $U$ is not a 
unitary matrix.

Since $U$ is the matrix which diagonalizes a symmetric matrix, it may be parametrized as
\begin{equation}
U=\left(\begin{array}{ccc}
c_{12}c_{13} & s_{12}c_{13} & s_{13}\\
-s_{12}c_{23}-c_{12}s_{13}s_{23} &c_{12}c_{23}-s_{12}s_{13}s_{23} &c_{13}s_{23} \\
s_{12}s_{23}-c_{12}s_{13}c_{23} &-c_{12}s_{23}-s_{12}s_{13}c_{23} &c_{13}c_{23} \\
\end{array}
\right),
\label{unu}
\end{equation}
where $c_{ij}=\cos\theta_{ij}$ and $s_{ij}=\sin\theta_{ij}e^{i\alpha_{ij}}$.
We can represent the neutrino Yukawa coupling $h_{\alpha k}$ and the mass eigenvalue $m_k$
by using this $U$ as eq.~(\ref{yukawa}) and eq.~(\ref{nmasseig}), respectively.

If $h_{\alpha 1}$ has the same structure as $h_{\alpha 2}$ for example, 
the mass matrix ${\cal M}_\nu$ can be diagonalized by $U$ given in eq.~(\ref{unu}) but it has
a zero mass eigenvalue $m_1=0$. On the formula of $m_2$ in eq.~(\ref{nmasseig}), 
$p_2h_2^2\Lambda_2$  is replaced by $p_1h_1^2\Lambda_1+p_2h_2^2\Lambda_2$. 
As such a concrete example, neutrino Yukawa couplings may be taken  
as \cite{zeromass}
\begin{equation} 
h_{e i}=h_{\mu i}=-h_{\tau i}=\frac{1}{\sqrt 3}h_i ~~(i=1,2), \quad
h_{e 3}=0, ~ h_{\mu 3}=h_{\tau 3}=\frac{h_3}{\sqrt 2}. 
\label{tribi}
\end{equation}
In that case, $U$ is found to be the tribimaximal mixing matrix \cite{otribi}  
and the mixing angles are fixed as 
\begin{equation}
s_{12}=\frac{1}{\sqrt 3}, \quad s_{23}=\frac{1}{\sqrt 2}, \quad s_{13}=0, 
\quad \alpha_{12}=\alpha_{23}=0, 
\end{equation}
and the neutrino mass eigenvalues are obtained as 
\begin{equation}
m_1=0,\quad m_2=h_1^2\Lambda_1+h_2^2\Lambda_2,  \quad   m_3=h_3^2\Lambda_3.
\end{equation}
The fault of the tribimaximal mixing matrix as the PMNS matrix could be overcome 
by considering a nontrivial flavor mixing matrix $U_e$ in the charged lepton sector \cite{zeromass}.
Since $U_e$ is expected to be almost diagonal, it can become a phenomenologically viable possibility. 

Here, we remark on an important consequence caused by the feature of this mass matrix 
addressed above.
If we consider leptogenesis based on the out-of-equilibrium decay of the right-handed neutrino
$N_k$, $CP$ asymmetry in this process is known to be caused by the interference between tree and 
one-loop diagrams and it should be nonzero at least to make leptogenesis work.
On this point, the model predicts an interesting result. 
Since this $CP$ asymmetry is proportional to 
${\rm Im}(\sum_\alpha h_{\alpha k}^\ast h_{\alpha j})^2$ for $k\not= j$ as 
shown in eqs.~(\ref{cp}) and (\ref{res}) later, this condition cannot be satisfied 
because of $\sum_\alpha h_{\alpha k}^\ast h_{\alpha j}\propto\delta_{kj}$ for the real neutrino 
Yukawa coupling $h_{\alpha k}$. This is the case even if Majorana phases in the right-handed 
neutrino mass are taken into account. Thus, the Yukawa coupling $h_{\alpha k}$ should be 
complex, and the PMNS matrix cannot be unitary in that case.
However, we should note that there is an exceptional case where the $CP$ asymmetry 
can take a nonzero value even if the Yukawa coupling $h_{\alpha k}$ is real. 
The existence of a zero mass eigenvalue presents such a case as found in an example given in
eq.~(\ref{tribi}) when Majorana phases exist in the right-handed neutrino mass \cite{zeromass}.
The PMNS matrix becomes unitary in this case.  

Neutrino oscillation data fix squared mass differences $\Delta m_{32}^2\equiv m_3^2-m_2^2$ 
and $\Delta m_{21}^2\equiv m_2^2-m_1^2$. 
If $h_1$ is sufficiently small and then $m_1\simeq0$, Yukawa couplings $h_2$ and 
$h_3$ can be determined by applying eq.~(\ref{nmasseig}) to
\begin{equation}
m_2=\sqrt{\Delta m_{21}^2}, \qquad m_3=\sqrt{\Delta m_{32}^2+\Delta m_{21}^2}.
\label{delm}
\end{equation}
If we use the best fit values from neutrino oscillation data given in \cite{pdg} for $\Delta m_{ij}^2$ and 
$\sin\theta_{ij}$ in the matrix $U$, we can calculate two effective masses relevant to two observables, 
that is, the effective masses for the neutrinoless double $\beta$ decay and the $\beta$ decay. 
They are predicted to be  
\begin{equation}
m_{\beta\beta}\simeq\left|\sum_{j=1}^3 U_{ej}^2m_j\right|= 0.0037~{\rm eV}, \quad
m_{\beta}\simeq\sqrt{\sum_{j=1}^3 |U_{ej}|^2m_j^2}=0.0088~{\rm eV}
\end{equation}
in normal ordering case.
These results are much smaller than their present bounds \cite{katrin,kamzen}. 

In the following part, we focus our study on a case where neutrino masses take normal ordering and
$N_2$ and $\eta_R$ have almost degenerate mass.\footnote{$\Lambda_k$ can be approximated as
$\Lambda_k=-\frac{\lambda_5\langle\phi\rangle^2}{64\pi^2M_k}$ in a case that $M_k$ 
and $M_{\eta_R}$ are almost degenerate.}  
Their masses are described as $M_{\eta_R}=(1-\delta_2)M_2$ with $\delta_2\ll 1$. 
To make the model clearer, we consider a numerical example of the relevant 
parameters which can derive $\Delta m_{32}^2$ and $\Delta m_{21}^2$ required by the neutrino 
oscillation data \cite{pdg}. As such an example for $M_k$, $M_{\eta_R}$ and Yukawa couplings, 
we adopt\footnote{Since DM is supposed to be $N_1$ here,  there is no constraint on a 
lower bound of $|\lambda_5|$ differently from the case where $\eta_R$ is 
DM \cite{ks}. A smaller $|\lambda_5|$ makes $h_k$ larger for fixed values of $M_k$ 
as found from eq.~(\ref{scot}). }
\begin{eqnarray}
&& M_{\eta_R} \simeq M_2\simeq M_3\simeq 3~{\rm TeV},  
\quad \lambda_5=- 10^{-3}, \nonumber \\
&&h_2=1.11\times 10^{-3}, \quad h_3=4.15\times 10^{-3},  
\label{bench}
\end{eqnarray}
where we assume $\alpha_{12}=\alpha_{13}=0,~\alpha_{23}=\frac{\pi}{2}$ and use for $\sin\theta_{ij}$
the best fit values from neutrino oscillation data.
$M_1$ and $h_1$ can be fixed freely as long as $h_1$ is small enough. 
We note that the present bounds of LFV like $\mu\rightarrow e\gamma$ \cite{pdg} 
can be satisfied for these parameters. We come back to this point in the last part of the paper again.

\subsection{Relic abundance of right-handed neutrino DM}
Right-handed neutrino $N_1$ is stable to be DM under the condition (\ref{massorder}).
However, its abundance is difficult to reach the required value in the freezeout scenario 
as mentioned in the introduction.
In fact, it can be produced in the thermal bath at the temperature higher than its mass through both the 
$\eta$ decay and the scatterings ($\eta^\dagger\eta\rightarrow N_1N_1^c$ and
$\ell_\alpha^\dagger\ell_\alpha\rightarrow N_1N_1^c$) caused by the neutrino Yukawa coupling 
$h_{\alpha 1}$ if it is large enough.
In that case, $h_{\alpha 1}$ has to take a value of $O(1)$ to reduce its abundance 
sufficiently through the annihilation. Unfortunately, such a large Yukawa coupling causes 
LFV processes like $\mu\rightarrow e\gamma$ at an unacceptable level 
through the one-loop diagram \cite{mks}. 
Since this difficulty is caused by a feature that $N_1$ has interaction due to the neutrino 
Yukawa couplings only, new interaction like $U(1)^\prime$ has been introduced to reduce 
its abundance in many articles \cite{comm,u1}.

In the present model, since the neutrino Yukawa coupling $h_1$ is supposed to be very small
as shown in eq.~(\ref{bench}), its thermal abundance is originally difficult to reach a value 
larger than the required one so that the freezeout scenario cannot be applied to the model. 
However, even in that case, we can consider the freezein scenario \cite{freezein} as an alternative possibility\footnote{A right-handed neutrino DM through the freezein mechanism has been 
studied in the different model \cite{n1dm} and  in the original 
scotogenic model \cite{nonext1,nonext2}.} in which $N_1$ is produced through the $\eta$ 
decay and the scatterings caused by the neutrino Yukawa coupling $h_1$. 
Since $h_1$ is supposed to be sufficiently small here, the $N_1$ generation through these 
processes is expected to be largely suppressed. 
On the other hand, the next lightest DM candidate 
$\eta_R$ is in the thermal equilibrium and decrease of its number density is controlled also by its 
coannihilation \cite{coan} caused by the SM interactions and the couplings $\lambda_{3,4,5}$ 
in eq.~(\ref{lag}).
After the freezeout of this coannihilation, the relic $\eta_R$ finally decays to $N_1$. 
Thus,  we have to take into account that there is a case where the contribution from 
such $N_1$s plays a crucial role in the estimation of the $N_1$ relic abundance.

We express the number density of $\eta_R$ and $N_1$ in the 
comoving volume by $Y_i=\frac{n_i}{s}$, where $n_i$ is the number density of particle 
species $i$ and $s$ is the entropy density.
If we define a dimensionless parameter $z_2$ as 
$z_2\equiv \frac{M_2}{T}$ and represent an equilibrium value of $Y_i$ as $Y_i^{\rm eq}$, 
Boltzmann equations for $Y_{\eta_R}$ and $Y_{N_1}$ 
can be written as 
\begin{eqnarray}
&&\frac{dY_{\eta_R}}{dz_2}=-\frac{z_2}{H(M_2)s}\left\{2 \gamma^S(\eta_R\eta_R)
+\gamma^D(\eta_R)\right\}\left(\frac{Y_{\eta_R}}{Y_{\eta_R}^{\rm eq}}-1\right),   \label{bolt1} \\
&&\frac{dY_{N_1}}{dz_2}=\frac{z_2}{H(M_2)s}\gamma^D(\eta_R)
\left(\frac{Y_{\eta_R}}{Y_{\eta_R}^{\rm eq}}-1\right),
\label{bolt2}
\end{eqnarray}
where $H(M_2)$ represents the Hubble parameter at $T=M_2$.
Reaction densities $\gamma^D(a)$ and $\gamma^S(ab)$ are respectively related to 
thermally averaged decay width $\langle\Gamma_a\rangle$ for $a\rightarrow ij$ and 
thermally averaged cross section $\langle\sigma_{ab}|v|\rangle$ for the 
scattering $ab\rightarrow ij$ as\cite{luty} 
\begin{equation}
\gamma^D(a)=\langle\Gamma_a\rangle n_a^{\rm eq}, \qquad
\gamma^S(ab)=\langle\sigma_{ab}|v|\rangle n_a^{\rm eq}n_b^{\rm eq},
\end{equation} 
where $n_a^{\rm eq}$ is the equilibrium number density of a particle $a$.
The present DM abundance can be 
expressed as $\Omega_{DM}=M_{DM}Y_{DM}^\infty s_0/\rho_0$, where
$\rho_0$ and $s_0$ are the energy density and the entropy density 
in the present Universe. Since they are given as $\rho_0=3M_{\rm pl}^2 H_0^2$ and 
$s_0=2.27\times 10^{-38}$~GeV$^3$, the required $N_1$ relic abundance $Y_{N_1}^\infty$ 
is found to be expressed as $Y_{N_1}^\infty=4.2\times 10^{-10}(M_1/{\rm GeV})^{-1}$ 
from $\Omega_{N_1}h^2=0.12$.

Here, it is instructive to give a rough explanation of how a $h_1$ value 
changes the role of the $\eta_R$ decay for the $N_1$ relic abundance. For that purpose, 
we estimate both the freezeout temperature $T_\eta$ of the coannihilation of $\eta_R$
and a value of $h_1$ which makes the $\eta_R$ decay reach the equilibrium at $T_\eta$.
Such a value of $h_1$ can be estimated from the condition $\Gamma_{\eta_R}=H(T_\eta)$.
The $\eta_R$ decay width $\Gamma_{\eta_R}$ and the Hubble parameter $H(T)$ are given as
\begin{equation}
\Gamma_{\eta_R}=\frac{q_1h_1^2}{32\pi}M_{\eta_R}\left(1-\frac{M_1^2}{M_{\eta_R}^2}\right)^2,
\quad H(T)=\left(\frac{\pi^2}{90}g_\ast\right)^{1/2}\frac{T^2}{M_{\rm pl}},
\label{hub}
\end{equation} 
where $M_{\rm pl}$ is the reduced Planck mass and $q_k$ is defined 
as $q_k=\sum_\alpha(U_{\alpha k}^\ast U_{\alpha k})$.
This condition determines $h_1$ as
\begin{equation}
h_1c_p=3\times 10^{-8}\left(\frac{M_{\eta_R}}{3~{\rm TeV}}\right)^{1/2},
\label{condh1}
\end{equation}
where $c_p$ is a constant determined by the phase space volume.\footnote{When the masses 
of $\eta_R$ and $N_1$ take closely degenerate values, $c_p$ can be represented as 
 $c_p=2\delta_1$ where $\delta_1$ is defined by $M_1=(1-\delta_1)M_{\eta_R}$ with $\delta_1\ll 1$.}
The freezeout temperature of the coannihilation $T_\eta(\equiv M_{\eta_R}/z_\eta)$ 
can be estimated through 
a condition $2n_{\eta_R}^{\rm eq}(T_\eta)\langle\sigma_{\eta\eta}|v|\rangle =H(T_\eta)z_\eta$ 
 as found from eq.~(\ref{bolt1}) by assuming instantaneous freezeout. 
Using the Hubble parameter $H(T)$ given in eq.~(\ref{hub}), $z_\eta$ can be determined by solving this 
condition as \cite{dmrelic}
\begin{equation}
z_\eta=\ln\frac{0.384 M_{\rm pl} \langle\sigma_{\eta\eta} |v|\rangle M_{\eta_R}}
{(g_\ast z_\eta)^{1/2}}.
\label{dtemp}
\end{equation}

If $h_1$ takes a larger value than the one given in eq.~(\ref{condh1}), 
the $N_1$ relic abundance $Y_{N_1}^\infty$ is expected 
to be approximated well as $Y_{N_1}^\infty=Y_{\eta_R}^{\rm eq}(T_c)$ through the freezein 
caused by the $\eta_R$ decay. $T_c$ is defined by 
$Y_{N_1}(T_c)=Y_{\eta_R}^{\rm eq}(T_c)$ since the $\eta_R$ decay at $T<T_c$ does not 
affect $Y_{N_1}^\infty$ substantially.  
$Y_{N_1}^\infty$ is determined only by the coupling $h_1$ for a fixed $M_1$ and
the freezeout of the $\eta_R$ coannihilation does not affect this result. 
However, this situation could change when $h_1$ takes a smaller value than the one 
given in eq.~(\ref{condh1}) and $T_c<T_\eta$ is satisfied.   
In this case, the $N_1$ relic abundance $Y_{N_1}^\infty$ is expected to be affected by 
the equilibrium number density of $\eta_R$ at $T_\eta$ since $\eta_R$ decays to $N_1$ finally. 
$T_\eta$ is determined through eq.~(\ref{dtemp}) 
by $\langle \sigma_{\eta\eta}|v|\rangle$ which depends on 
the couplings $\lambda_{3,4}$ and $M_{\eta_R}$.
In the study of the inert doublet DM model \cite{inert}, these couplings are 
known to play a crucial role there. 
If we take account of it, the $N_1$ relic abundance  
could be predicted through a relation $Y_{N_1}^\infty \simeq Y_{\eta_R}^{\rm eq}(T_\eta)$.
By imposing the DM relic abundance on $N_1$, 
we can find constraints on the couplings $\lambda_{3,4}$ through a condition
\begin{equation}
Y_{\eta_R}^{\rm eq}(T_\eta)\simeq 4.2\times 10^{-10}\left(\frac{1~{\rm GeV}}{M_1}\right).
\label{yetaeq}
\end{equation}
If the coannihilation occurs only through the SM interactions ($\lambda_{3,4}=0$), 
we find $Y_{\eta_R}^{\rm eq}(T_\eta)=4.9\times 10^{-11}$. Since  
$Y_{\eta_R}^{\rm eq}(T_\eta)$ decreases for $|\lambda_{3,4}|\not=0$, 
 eq.~(\ref{yetaeq}) shows that $M_1~{^>_\sim}~8.5$ GeV should be satisfied to obtain 
the suitable $N_1$ abundance.

\begin{figure}[t]
\begin{center}
\includegraphics[width=8.5cm]{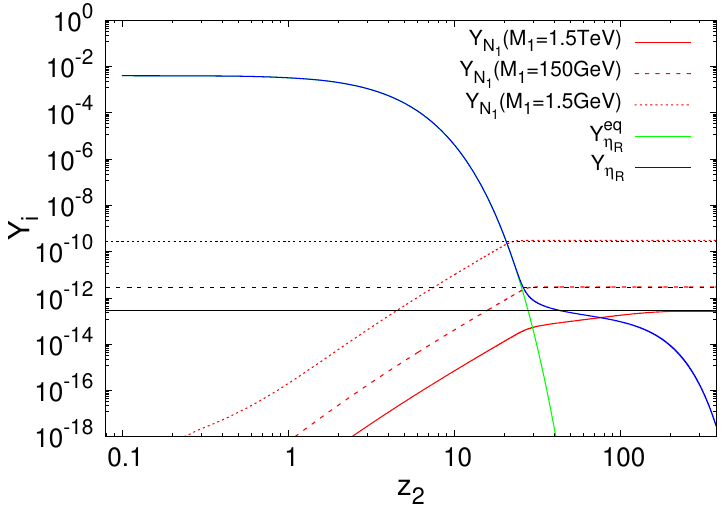}
\hspace*{5mm}
\includegraphics[width=6.5cm]{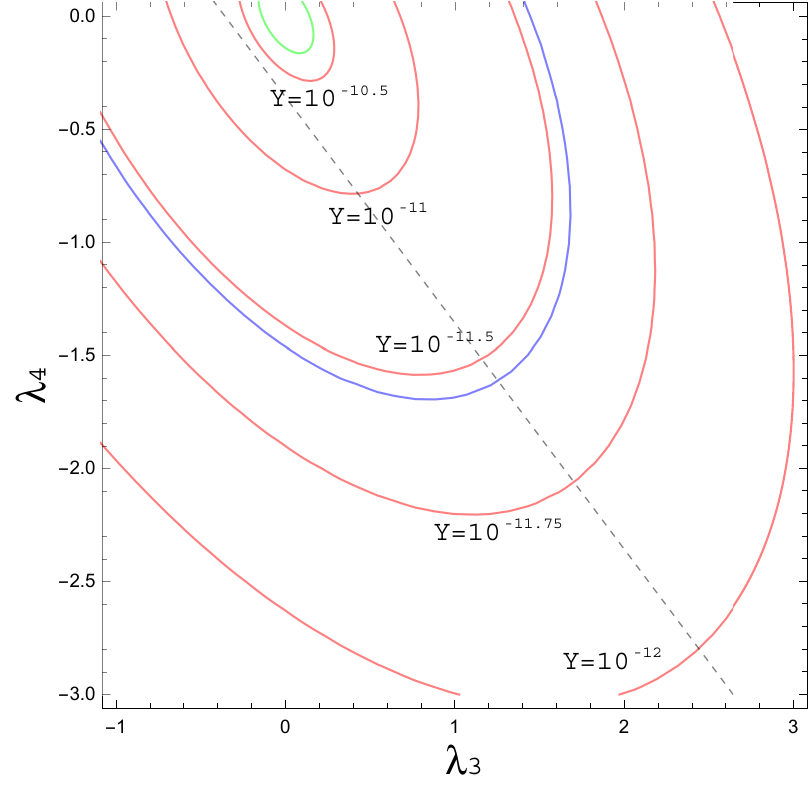}
\end{center}

\footnotesize{{\bf Fig.~1}~~~Left panel:~Evolution of the number density of $N_1$ and $\eta_R$ 
in the comoving volume.  They are plotted with red lines for three values of $M_1$,
1.5 TeV (solid), 150 GeV (dashed) and 1.5 GeV (dotted). $h_1$ is fixed 
at $7.8\times 10^{-9}$, $4.5\times 10^{-8}$ and $7.0\times 10^{-6}$ 
for each $M_1$, respectively. $M_{\eta_R}=3$~TeV and $(\lambda_3,\lambda_4)=(1.8,-1.0)$ 
are assumed. Horizontal black lines represent $Y_{N_1}^\infty$ values $2.8\times 10^{-13}$,
$2.8\times 10^{-12}$, $2.8\times 10^{-10}$ required 
to realize the present DM relic density for each value of $M_1$. 
While $Y_{\eta_R}$ for $M_1=1.5$ TeV deviates from
$Y_{\eta_R}^{\rm eq}$,  $Y_{\eta_R}$ in other two cases coincides with $Y_{\eta_R}^{\rm eq}$ 
 since the $\eta_R$ decay is already in the equilibrium at a temperature higher than $T_\eta$.
Right panel:~Contours of $Y_{\eta_R}^{\rm eq}(T_\eta)$ for $M_{\eta_R}=3$~TeV and $h_1=0$ 
in the $(\lambda_3,\lambda_4)$ 
plane.  Each red line represents $Y_{\eta_R}^{\rm eq}(T_\eta)=10^{-10.5}, 10^{-11}, 10^{-11.5}, 10^{-11.75}$, 
and $10^{-12}$ outward. Green and blue lines represent $Y_{\eta_R}^{\rm eq}(T_\eta)$ 
given in eq.~(\ref{yetaeq}) for $M_1=10$ GeV and 150 GeV. 
A dashed line represents $\lambda_3+\lambda_4+\lambda_5=-\sqrt{\lambda_1\lambda_2}$ in the case of 
$\lambda_2=1.0$, and an upper region of it satisfies the potential stability. 
No constraint comes from direct detection experiments on $\lambda_{3,4}$ since DM is $N_1$.}
\end{figure}

In order to show this observation concretely, we solve the Boltzmann equations for 
$Y_{N_1}$ and $Y_{\eta_R}$ using the parameters given in eq.~(\ref{bench}) and 
$(\lambda_3,\lambda_4)=(1.8, -1.0)$.
Since $M_1$ and $h_1$ can be free from neutrino mass determination as addressed 
below eq.~(\ref{bench}), we plot the solutions in the left panel for the cases 
$(M_1,h_1)=(1.5~{\rm TeV}, 7.8\times 10^{-9})$, $(150~{\rm GeV}, 4.5\times 10^{-8})$, 
and $(1.5~{\rm GeV}, 7.0\times 10^{-6})$ where $h_1$ is tuned to give the required DM abundance.
In this panel, $Y_{N_1}^\infty$ required to explain the DM abundance is shown by horizontal black 
lines for each case.

In the latter two cases with small $M_1$, $T_c>T_\eta$ is satisfied and the required $Y_{N_1}^\infty$ is 
found to be realized by the pure freezein due to the decay of $\eta_R$ 
in the thermal equilibrium through a relation 
$Y_{N_1}^\infty=Y_{N_1}(T_c)=Y_{\eta_R}^{\rm eq}(T_c)$.  
$Y_{\eta_R}$ does not deviate from the equilibrium one for the assumed values 
of $h_1$ and $\lambda_{3,4}$ there.  
In the right panel,  contours of  $Y_{\eta_R}(T_\eta)$ in the case $M_{\eta_R}=3$~TeV
are plotted for $Y_{\eta_R}^{\rm eq}(T_\eta)=
10^{-10.5}, 10^{-11}, 10^{-11.5}, 10^{-11.75}$, and $10^{-12}$ with red lines
in the $(\lambda_3,\lambda_4)$ plane.
This figure shows that $Y_{\eta_R}^{\rm eq}(T_\eta)$ is much smaller to realize the required 
$Y_{N_1}^\infty$ for $(\lambda_3,\lambda_4)=(1.8,-1.0)$ which is assumed in the left panel.
The freezeout of the $\eta_R$ coannihilation is irrelevant to the estimation of $Y_{N_1}^\infty$ 
in these cases.

As an extreme case for this kind of small $N_1$ examples, we may consider $N_1$ with keV scale mass.
We can confirm that the required relic DM abundance is obtained by tuning $h_1$ suitably,    
for example, $h_1=3.7\times 10^{-3}$ for $M_1=5$~keV. 
We note that the values of $h_{2,3}$ in eq.~(\ref{bench}) are not affected even if $h_1$ takes 
the same order value as $h_{2,3}$ as long as $M_1$ is much smaller than $M_{\eta_R}$ as 
in the present case. 
It is caused by the fact that $\Lambda_1$ has a strong suppression factor of $O(M_1/M_{\eta_R})$ 
compared with $\Lambda_{2,3}$. Thus, $N_1$ with keV scale mass is allowed to have an 
acceptable relic abundance as DM for $h_{2,3}$ given in eq.~(\ref{bench}) without 
contradicting with the neutrino oscillation data. 

In the case of $M_1=1.5$ TeV,  the freezeout of the $\eta_R$ coannihilation is 
relevant to the estimation of $Y_{N_1}$.  
Although $Y_{N_1}(T_c)$ is smaller than the required value $Y_{N_1}^\infty$,
the late time decay of the freezeout $\eta_R$ contributes 
to $Y_{N_1}$ to increase it to $Y_{N_1}^\infty$. 
In fact, the required $Y_{N_1}^\infty$ can be obtained as shown in the left 
panel if $\lambda_3$ and $\lambda_4$ are fixed to suitable values in a part 
over the dashed line in the $(\lambda_3,\lambda_4)$ 
plane shown in the right panel, where the stability condition (\ref{stab}) of the potential 
$V$ in eq.~(\ref{lag}) is satisfied.

In the above study, we fix the couplings $\lambda_{3,4}$ and $h_1$ 
is tuned to realize the required $Y_{N_1}^\infty$. However, if we suppose that the decay of 
the freezeout $\eta_R$ is the dominant origin of $Y_{N_1}^\infty$ as 
in the $M_1=1.5$ TeV case, we could find the bound on $\lambda_{3,4}$ from the DM study. 
Here, we consider such a possibility in the two small $M_1$ 
cases assuming that the $\eta_R$ decay caused by $h_1$ is negligibly small 
and $Y_{\eta_R}^{\rm eq}$ is kept
only by the SM interactions and the $\lambda_{3,4}$ couplings. In that case,
 the required $Y_{\eta_R}^\infty$ is found to be $2.8 \times 10^{-10}$ 
and $2.8\times 10^{-12}$ for $M_1=1.5$ GeV and 150 GeV by using eq.~(\ref{yetaeq}). 
If we recall the result given below eq.~(\ref{yetaeq}), the suitable $Y_{N_1}^\infty$ cannot be obtained 
without help of the decay of $\eta_R$ in the equilibrium caused by $h_1$ in the former case. 
On the other hand, in the latter case, it can be obtained for suitable values of $\lambda_{3,4}$ 
without taking account of such early time $\eta_R$ decay.
$Y_{\eta_R}^{\rm eq}(T_\eta)$ in that case is plotted by a blue line in the right 
panel,\footnote{In this panel, $M_1=10$ GeV case is also plotted 
by a green line as a reference.} 
from which we can read off the required values of $\lambda_{3,4}$.
Generally, if the $h_1$ value increases, the required $Y_{\eta_R}^\infty$ decreases and
the couplings $|\lambda_{3,4}|$ take larger values. It is confirmed from this example.
This behavior is brought about from the $N_1$ contribution caused from the early time 
decay of $\eta_R$ in the equilibrium. 
It should be also noted that there is no constraint on the values of  $\lambda_3$ and $\lambda_4$
from direct detection experiments since $N_1$ is DM. 
It is a crucial difference from the case where $\eta_R$ is DM.
The coupling constants $\lambda_3$ and $\lambda_4$ could be also constrained through 
the correction to the Higgs mass caused by them. This constraint is reported  in \cite{mhiggs} 
that it is satisfied for $M_{\eta_R} \le 4.4$ TeV. 

In the above study, we do not take into account the effect caused by the simultaneous explanation 
of the baryon number asymmetry in the Universe through leptogenesis which is
 based on the $N_2$ decay.
Since it could produce $\eta_R$ at late time, the $N_1$ abundance could be 
affected by it if the $N_2$ decay 
occurs near the freezeout of the coannihilation of $\eta_R$. 
We discuss this point in the next part.

\section{Consistency with leptogenesis}
In the present model, the baryon number asymmetry is considered to be produced 
through leptogenesis caused by the out-of-equilibrium decay of $N_2$. 
This requires that $N_2$ should be in the thermal bath and its decay to $\eta_R$ occurs in a $CP$ asymmetric way at a higher temperature than 100 GeV where sphaleron interaction 
is in the equilibrium. 
Values of $h_{2,3}$ given in eq.~(\ref{bench})  are large enough to produce $N_{2,3}$ 
in the thermal bath as relativistic particles through the inverse decay and scatterings 
caused by them as long as reheating temperature is larger than $M_3$.
The $CP$ asymmetry in the decay $N_2\rightarrow \ell_\alpha\eta^\dagger$ can be 
caused through an interference between a tree diagram and a one-loop diagram mediated 
by $N_1$ and $N_3$.

This $CP$ asymmetry is estimated through the formula  
\begin{eqnarray}
&&\varepsilon=\frac{1}{8\pi}\sum_{k=1,3}\frac{ {\rm Im}
\left[\left(\sum_\alpha h_{\alpha 2}h^\ast_{\alpha k}\right)^2\right]}
{ \sum_\alpha h_{\alpha 2}h_{\alpha 2}^\ast}
F\left(\frac{M_k^2}{M_2^2},\frac{M_{\eta_R}^2}{M_2^2}\right)\nonumber \\
&&F(x,y)=\sqrt{x}\left[1-\frac{x-2y+1}{(1-y)^2}\ln\left(\frac{x-2y+1}{x-y^2}\right)\right],
\label{cp} 
\end{eqnarray}
where $M_2\simeq M_{\eta_R}$ is taken into account.
Since $h_1$ is supposed to be very small for the DM explanation 
as discussed in the previous part, 
the dominant contribution to this $\varepsilon$ is found to be caused by $N_3$ and 
it can be estimated by using eqs.~(\ref{nmasseig}) and (\ref{delm})
as\footnote{As addressed in the previous part, it should be noted that the existence of both 
a zero mass eigenvalue and Majorana phases of the right-handed neutrino masses 
is required to guarantee a non-zero $\varepsilon$ when neutrino Yukawa couplings 
$h_{\alpha k}$ are real \cite{zeromass}.} 
\begin{eqnarray}
\varepsilon&\simeq&
\frac{h_3^2}{8\pi}
\frac{{\rm Im}[(\sum_\alpha U_{\alpha 2}U_{\alpha 3}^\ast)^2e^{i(\gamma_2-\gamma_3)}]}
{\sum_\alpha U_{\alpha 2}U_{\alpha 2}^\ast}
F\left(\frac{M_3^2}{M_2^2}, \frac{M_{\eta_R}^2}{M_2^2} \right)\nonumber\\
&\simeq& 4.7\times 10^{-7}F\left(\frac{M_3^2}{M_2^2}, \frac{M_{\eta_R}^2}{M_2^2} \right),
\label{e3}
\end{eqnarray}
where parameters in eq.~(\ref{bench}) are used and values of $\sin\theta_{ij}$ and $\alpha_{ij}$
are fixed to the ones written below eq.~(\ref{bench}).
In the second line, $\sin (\gamma_2-\gamma_3)=1$ is assumed.
If a smaller $|\lambda_5|$ is assumed in eq.~(\ref{bench}),  Yukawa couplings $h_{2,3}$ take larger 
values. In that case, $\varepsilon$ takes a larger value but the out-of-equilibrium 
decay of $N_2$ becomes harder. 

The $N_2$ decay at temperature $T\ge T_D$ is in the equilibrium if $T_D$ is defined by 
$\Gamma_{N_2}=H(T_D)$. The $N_2$ decay width $\Gamma_{N_2}$ can be expressed as
\begin{equation}
\Gamma_{N_2}=\frac{q_2h^2_2}{16\pi}M_2\left(1-\frac{M_{\eta_R}^2}{M_2^2}\right)^2.
\label{n2decay}
\end{equation}
If the washout processes of the generated lepton number asymmetry are frozen out at 
temperature $T_F$, the lepton number asymmetry caused by the out-of-equilibrium decay of 
$N_2$ could escape the washout in the case $T_D<T_F$. 
Such a situation could happen if the $N_2$ decay is forced to be delayed by making
$\Gamma_{N_2}$ smaller. 

Although $h_2$ fixed by the neutrino oscillation data cannot realize such a situation 
for a TeV scale $M_2$,
it could be realized through an additional suppression factor for $\Gamma_{N_2}$. 
It is caused by the phase space suppression due to the strict degeneracy between 
$M_2$ and $M_{\eta_R}$ \cite{ds}.
In fact, if their degeneracy is expressed as $M_{\eta_R}=(1-\delta_2)M_2$ with $\delta_2\ll 1$,
$\Gamma_{N_2}$ is found to be rewritten with an additional suppression factor $\delta_2$ as 
\begin{equation}
\Gamma_{N_2}\simeq \frac{h_2^2\delta_2^2}{4\pi}M_2
= 2.9\times 10^{-12}\left(\frac{M_2}{3~{\rm TeV}}\right)\left(\frac{\delta_2}{10^{-4}}\right)^2~{\rm GeV}.
\label{gam2}
\end{equation}
If we assume $\delta_2< 10^{-4}$ at $M_2=3$ TeV,
$\Gamma_{N_2}<H(T)$ is found to be satisfied for $T~{^<_\sim}~M_2$ and 
$N_2$ is expected to be out-of-equilibrium there. 
Thus, if the condition $T_D<T_F$ is fulfilled, 
the generated lepton number asymmetry $Y_L$ defined as $Y_L\equiv Y_\ell-Y_{\bar\ell}$ could 
be estimated as $Y_L=\varepsilon Y_{N_2}^{\rm eq}$ 
where $Y_{N_2}^{\rm eq}$ stands for the equilibrium value of $N_2$ as the relativistic particle. 
In that case, lepton number asymmetry $Y_L=O(10^{-10})$ required for the explanation of baryon 
number asymmetry in the Universe is expected to be obtained 
if $\varepsilon$ takes a value of $O(10^{-7})$ in eq.~(\ref{e3}).
Unfortunately, since $\delta_2\ll 1$ also suppresses $\varepsilon$ heavily as found 
from eq.(\ref{cp}), this idea cannot work in the present case.

If $M_2\simeq M_1$ or $M_3$ is satisfied, the $CP$ asymmetry $\varepsilon$ could be enhanced by 
the resonance due to a self-energy diagram mediated by
$N_1$ or $N_3$. It is expressed as  \cite{cpasym0,res,cpasym}
\begin{eqnarray}
\varepsilon&=&\frac{ {\rm Im}
\left[\left(\sum_\alpha h_{\alpha 2}h^\ast_{\alpha k}\right)^2\right]}
{ (\sum_\alpha h_{\alpha 2}h_{\alpha 2}^\ast)
(\sum_\alpha h_{\alpha k}h_{\alpha k}^\ast)}
\frac{(M_2^2-M_k^2)M_2\Gamma_{N_k}}{(M_2^2-M_k^2)^2+M_2^2\Gamma_{N_k}^2},
\label{res}
\end{eqnarray} 
where $k=1, 3$ and $\Gamma_{N_k}$ is the decay width of $N_k$.
Since $N_1$ is stable in the present model and then $\Gamma_{N_1}=0$, 
only $N_3$ can contribute to the resonant $CP$ asymmetry. Thus, 
$\varepsilon$ can be written by using the values for $\sin\theta_{ij}$ and 
$\alpha_{ij}$ given below eq.~(\ref{bench})  as
\begin{equation}
\varepsilon= -{\rm Im}
\left[\left(\sum_\alpha U_{\alpha 2}U^\ast_{\alpha 3}\right)^2 e^{i(\gamma_2-\gamma_3)}\right]
\frac{2\delta_3\tilde\Gamma_{N_3}}{4\delta_3^2+\tilde\Gamma_{N_3}^2}=
0.67\frac{2\delta_3\tilde\Gamma_{N_3}}{4\delta_3^2+\tilde\Gamma_{N_3}^2},
\end{equation}
where $M_3=(1+\delta_3)M_2$ and $\tilde\Gamma_{N_3}=\Gamma_{N_3}/M_2$, and 
$\sin (\gamma_2-\gamma_3)=1$ is assumed.
If $\delta_2>\delta_3$ is satisfied, $\tilde\Gamma_{N_3}$ can be estimated as 
$\tilde\Gamma_{N_3}=h_3^2\delta_2^2/4\pi$ and then we have
$\varepsilon\simeq 0.027h_3^2\delta_2^2/\delta_3$ for $\tilde\Gamma_{N_3}<2\delta_3$.
Any special symmetry is not introduced to explain this strict mass 
degeneracy between $N_2$ and $N_3$. Since it is just assumed here,
quantum correction for the right-handed neutrino masses is caused at two-loop level. 
However, it can be confirmed to be sufficiently small. It is estimated as
\begin{equation}
\delta M_{ij}\simeq \sum_{\alpha,\beta}\frac{\lambda_5\langle\phi\rangle^2}{16\pi^2M_\eta}
h_{\alpha i}h_{\beta j}({\cal M}_\nu)_{\alpha\beta}.
\label{deltam}
\end{equation}
If we use the present parameters to estimate it, we find $\delta M_2=O(10^{-24})$ GeV
and $\delta M_3=O(10^{-23})$ GeV.\footnote{It is caused by the smallness 
of $|\lambda_5|$ which is justified by the lepton number. If the lepton number is assigned to $\eta$, 
we find that it is violated only through the $\lambda_5$ term.}  
They are small enough not to disturb the assumed mass 
difference between $N_2$ and $N_3$.

It is crucial to know  what value of $\delta_2$ can make the condition 
$100~{\rm GeV}<T_D<T_F$ be satisfied for the parameters given in eq.~(\ref{bench}).
By calculating the relevant reaction rates of washout processes, 
we find that the freezeout of the $\eta_R$ coannihilation occurs for $T_F~{^>_\sim}~0.1M_2$.
Thus, $T_D<T_F$ is guaranteed if $\delta_2<3\times 10^{-5}$ is satisfied. 
This suggests that the $CP$ asymmetry $\varepsilon$ can take a value of $O(10^{-7})$
in the case $\delta_3=O(10^{-9})$, which makes successful leptogenesis possible 
as addressed below eq.~(\ref{gam2}).
However, if the number density of $N_2$ does not decrease to the required $Y_{N_1}^\infty$
at the temperature where the $\eta_R$ coannihilation freezes out, 
the $\eta_R$ produced through the $N_2$ decay could contribute to the $N_1$ relic density.
Thus, the estimation of the $N_1$ relic abundance in the previous section
cannot be naively applied to this scheme for leptogenesis.
The effect of $\eta_R$ produced in the late time $N_2$ decay needs to be taken into 
account in the estimation of the $N_1$ relic abundance. 

In order to examine its effect to the $N_1$ relic abundance,  
we solve numerically Boltzmann equations for $Y_{N_2}$, $Y_L$,  $Y_{\eta_R}$ and $Y_{N_1}$.
Using a dimensionless parameter $z_2$, they are given as
\begin{eqnarray}
\frac{dY_{N_2}}{dz_2}&=&-\frac{ z_2}{H(M_2)s}\left[\Big\{\gamma^S(N_2\ell) 
+\gamma^S(N_2\eta)\Big\}\left(\frac{Y_{N_2}^2}{Y_{N_2}^{{\rm eq}2}}-1\right) 
+\gamma^D(N_2)\left(\frac{Y_{N_2}}{Y_{N_2}^{\rm eq}}-1\right)\right], 
\nonumber \\
\frac{dY_L}{dz_2}&=&-\frac{z_2}{H(M_2)s}\left[-\varepsilon \gamma^D(N_2)
\left(\frac{Y_{N_2}}{Y_{N_2}^{\rm eq}}-1\right)
+\left\{\sum_{k=2,3}\gamma^D(N_k)+ 2(\gamma^S(\ell \eta^\dagger)
+\gamma^S(\ell\ell)) \right\}\frac{Y_L}{Y_\ell^{\rm eq}}\right], \nonumber \\
\frac{dY_{\eta_R}}{dz_2}&=&-\frac{z_2}{H(M_2)s}\left[\Big\{2 \gamma^S(\eta_R\eta_R)
+\gamma^D(\eta_R)\Big\}\left(\frac{Y_{\eta_R}}{Y_{\eta_R}^{\rm eq}}-1\right)
-\gamma^D(N_2)\left(\frac{Y_{N_2}}{Y_{N_2}^{\rm eq}}-1\right)\right],
\nonumber \\
\frac{dY_{N_1}}{dz_2}&=&\frac{z_2}{H(M_2)s}\gamma^D(\eta_R)
\left(\frac{Y_{\eta_R}}{Y_{\eta_R}^{\rm eq}}-1\right),
\label{bolt}
\end{eqnarray} 
where $\gamma^S({N_2\ell})$ and $\gamma^S(N_2\eta)$ are reaction densities of the scattering 
$N_3\ell\leftrightarrow N_2\ell$ and $N_3\eta\leftrightarrow N_2\eta$ for the $N_2$ production, 
and $\gamma^S(\ell\eta^\dagger)$ and $\gamma^S(\ell\ell)$ are the ones for the lepton number 
violating scattering $\ell\eta^\dagger \leftrightarrow \bar\ell\eta$ and 
$\ell\ell\leftrightarrow \eta\eta$.
A new term is introduced to the equation for $Y_{\eta_R}$ to take account of the contribution from 
the late time decay of $N_2$ in comparison with the original one in eq.~(\ref{bolt1}).
Change of the $Y_{N_1}$ evolution caused by imposing the production of the required lepton number 
asymmetry can be found from these solutions. 
Here, we should note that the simplified Boltzmann equations cannot be used 
for the leptogenesis if oscillation occurs between $N_2$ and $N_3$ during the leptogenesis. 
Although the right-handed neutrino masses are assumed to be flavor diagonal at tree level,
loop correction causes the off-diagonal mass as shown in eq.~(\ref{deltam}). 
However, it is much smaller than the assumed masses of $N_2$ and $N_3$ as addressed previously. 
Since the mixing between $N_2$ and $N_3$ is negligibly small,
the oscillation between $N_2$ and $N_3$ is found to be negligible and then these 
equations can be safely used to study the leptogenesis. 

Their solutions for the same values of $M_1$ and $(\lambda_3,\lambda_4)$ used in Fig.~1
are plotted as functions of $z_2$ in Fig.~2. 
We use the values of  $M_{\eta_R}$, $M_2$ and $M_3$ given in (\ref{bench}) with 
$\delta_2=3\times 10^{-5}$ and $\delta_3=10^{-9}$ which are adopted referring to the above 
discussion. 
The figure shows that the lepton number asymmetry $|Y_L|$ plotted by an orange line converges 
to a final value at $z_2\sim 10$. It is larger compared with a required 
range to explain the baryon number asymmetry shown by dotted yellow lines.
Since the resulting lepton number asymmetry takes a value 
$Y_L\simeq \varepsilon Y_{N_2}^{\rm eq}$ where $Y_{N_2}^{\rm eq}$ is the one of the relativistic particles, 
the washout of the generated lepton number asymmetry is found to be sufficiently suppressed
 as a result of the late time decay of $N_2$. It is expected 
in the above discussion.  

\begin{figure}[t]
\begin{center}
\includegraphics[width=8.5cm]{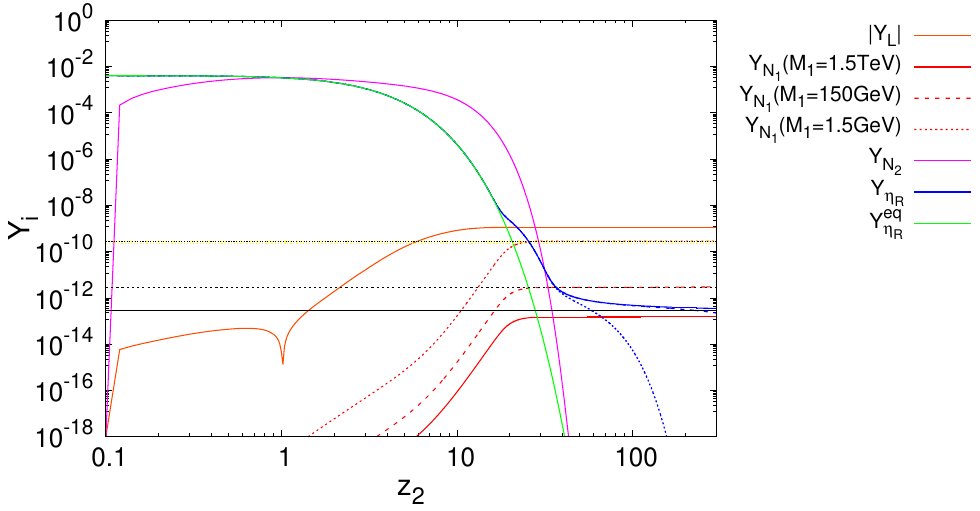}
\end{center}
\vspace*{-3mm}
\footnotesize{{\bf Fig.~2}~~Evolution of the lepton number asymmetry $|Y_L|$ 
and the $N_1$ number density $Y_{N_1}$ in the comoving volume. 
$|Y_L|$ is plotted by an orange line, which is caused by the out-of-equilibrium decay of $N_2$. 
$Y_{N_2}^{\rm eq}$ coincides with $Y_{\eta_R}^{\rm eq}$ since 
$M_2\simeq M_{\eta_R}$ is assumed. Yellow dashed horizontal lines represent a region 
of $|Y_L|$ to explain the required baryon number asymmetry.  
In the panel, $Y_{N_1}$ is plotted by red lines for the same $M_1$ as the ones 
in the left panel of Fig.~1. Only $h_1$ is tuned to realize the present DM abundance
which is represented by black solid, dashed and dotted lines 
for each $M_1$. Other parameters are fixed to the same values as the ones in Fig.~1.}
\end{figure}

\begin{figure}[t]
\begin{center}
\includegraphics[width=8.5cm]{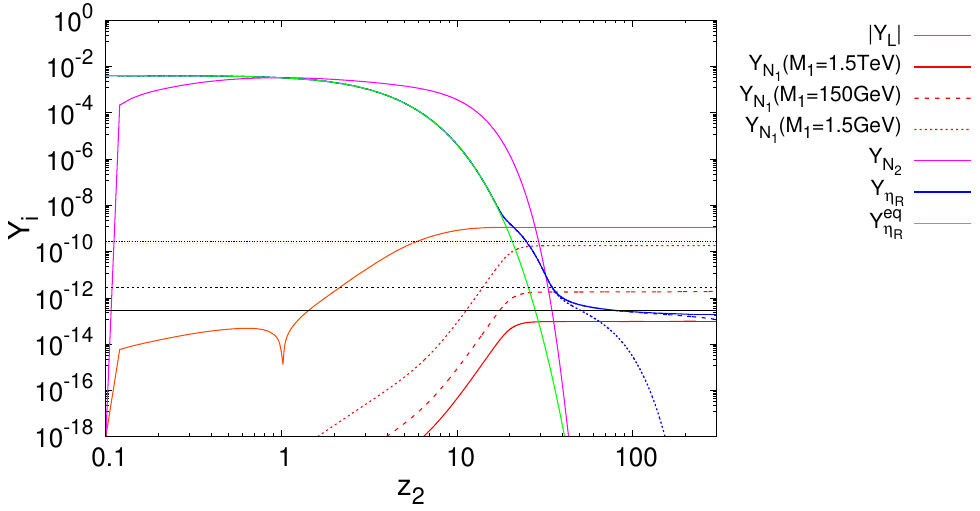}
\end{center}
\vspace*{-3mm}
\footnotesize{{\bf Fig.~3}~~Evolution of the lepton number asymmetry $|Y_L|$ 
and the $N_1$ number density $Y_{N_1}$ in the comoving volume. 
 $(\lambda_3,\lambda_4)$ is changed to $(2.5, -1.0)$ from the ones 
used in Fig.~2 keeping the $h_1$ value unchanged. 
While $Y_{N_1}$ becomes a little bit smaller than the required value $Y_{N_1}^\infty$ 
in the cases $M_1=150$ GeV and 1.5 GeV, 
$Y_{N_1}+Y_{\eta_R}=Y_{N_1}^\infty$ is realized for $M_1=1.5$ TeV at $z_2=300$.
It means that $Y_{N_1}=Y_{N_1}^\infty$ is obtained finally through the $\eta_R$ decay.  }
\end{figure}

In Fig.~2, $Y_{\eta_R}$ and $Y_{N_1}$ are plotted by using blue and red lines, respectively. 
Different line types are used for different masses, $M_1$=1.5 TeV (solid),  150 GeV (dashed), and
1.5 GeV (dotted). A $Y_{N_1}^\infty$ value which can reproduce $\Omega_{N_1} h^2=0.12$ as 
the present DM abundance is shown by horizontal black solid 
($Y_{N_1}^\infty=2.8\times 10^{-13}$), 
dashed ($Y_{N_1}^\infty=2.8\times 10^{-12}$) and dotted ($Y_{N_1}^\infty=2.8\times 10^{-10}$) lines for each $M_1$. 
We find from the figure that $Y_{\eta_R}$ leaves $Y_{\eta_R}^{\rm eq}$
around $z_2=20$ in these cases. It is considered to be caused due to the break of the balance 
between the $\eta_R$ production by the $N_2$ decay and its disappearance 
due to both the decay and the coannihilation. 
It shows that the late time additional $\eta_R$ production through the $N_2$
decay affects the estimation of the $N_1$ relic abundance. 
It forces us to change the strength of the $\eta_R$ decay and the $\eta_R$ coannihilation 
from the ones in Fig.~1 
in order to realize the required $N_1$ relic abundance\footnote{Since the $N_2$ decay 
is relevant to the leptogenesis, we do not change the parameters relevant to it here.}  
since we impose on the model the simultaneous 
explanation of the DM abundance and the baryon number asymmetry.

Since $h_1$ does not affect the values of $h_{2,3}$ determined by the neutrino 
oscillation data as long as $h_1$ is small enough, it can be tuned to result in the 
required $N_1$ relic abundance $Y_{N_1}^\infty$.
By tuning $h_1$ suitably to $h_1=2.3\times 10^{-9}$ and $2.3\times 10^{-8}$ for 
the case $M_1=$150 GeV and 1.5 GeV respectively, 
$Y_{N_1}$ is found to converge to the required $Y_{N_1}^\infty$ at $z_2$ 
where $Y_{N_1}=Y_{N_2}$ is realized.
These values of $h_1$ are much smaller than the ones used in Fig.~1.
It is considered to be caused to reduce the effect of $\eta_R$ which is additionally produced through the
$N_2$ decay by making the relative strength of the $\eta_R$ coannihilation to the $\eta_R$ 
decay larger. The similar feature can be observed in the evolution of the number density of $N_1$
with the keV scale mass also. For example, the DM relic abundance requires $h_1=6.5\times 10^{-5}$ for 
$M_1=5$ keV, which can be confirmed not to affect the result of leptogenesis.  
 In the case $M_1=1.5$ TeV, we find that the relation $Y_{N_1}+Y_{\eta_R}>Y_{N_1}^\infty$ cannot 
be changed by tuning  the value of $h_1$. 
This can be understood if we note $Y_{\eta_R}>Y_{N_1}^\infty$ at $z_2=300$
where the $\eta_R$ coannihilation is frozen out. 
It suggests that the freezeout of the $\eta_R$ coannihilation plays 
an important role and then $\lambda_3$ and $\lambda_4$ have to be tuned 
when $M_1$ takes a similar order value to $M_{\eta_R}$. 

In Fig.~3, $Y_{N_1}$ and $Y_{\eta_R}$ are plotted for the case 
where $\lambda_3$ and $\lambda_4$ are 
shifted from the ones in Fig.~2 to $(\lambda_3,\lambda_4)=(2.5, -1.0)$ keeping other 
parameters unchanged.
While $Y_{N_1}$ in the case of $M_1=150$ GeV and 1.5 GeV is found to take a little bit smaller 
value than the one in Fig.~2, $Y_{N_1}+Y_{\eta_R}$ converges to $Y_{N_1}^\infty$ 
at $z_2=300$ in the case $M_1=1.5$ TeV.
Since the $\eta_R$ coannihilation becomes stronger due to a larger value of $\lambda_3$, 
the contribution from $\eta_R$ is considered to be reduced.

These examples show that $N_1$ in a wide mass range can explain the required relic abundance 
of DM in a consistent way with successful leptogenesis by tuning $h_1$ and $\lambda_{3,4}$ suitably.
Couplings $\lambda_3$ and $\lambda_4$ play a crucial role especially when $M_1$ and 
$M_{\eta_R}$ take the same order values since $Y_{N_1}^\infty~{^>_\sim}~Y_{\eta_R}^\infty$ should be 
satisfied in that case. 
The model presents an interesting possibility 
such that three unsolved problems in the SM can be explained by low scale right-handed neutrinos 
simultaneously.

Finally, we should address phenomenological signatures of the model. 
Extra charged and neutral scalars $\eta^\pm$ and $\eta_{R,I}$ might be directly 
examined in the future collider experiments \cite{collider}.  
The collider signatures of $\eta$ in the present model are the same as 
the inert doublet model in the high mass region since the Yukawa coupling $h_1$ is very small.
However, the relevant parameter space can be extended since $\eta_R$ is not DM and
the constraints from the relic abundance and the direct detection are not applied.
As its results, there may be some processes that allow us to distinguish the present model from
the inert doublet model. 
Interesting ones are the production of $\eta^+\eta^-$, $\eta^\pm\eta_R$, $\eta^\pm\eta_I$, 
$\eta_R\eta_I$ whose signal can be observed as multi leptons + missing transverse energy and so on, 
and the precise measurements of the Higgs decay to $\gamma\gamma$ and $\gamma Z$.    
In the former case, since the decay width of $\eta^\pm$ is of $O(10^{-14})$ GeV for $h_1=O(10^{-8})$ 
and $|\lambda_4|=O(10^{-2})$,\footnote{We note that the decay width of $\eta^\pm$ mediated by 
a $W$ boson is proportional to $\Delta m^5$ where $\Delta m$ is the mass difference 
of $\eta^\pm$ and $\eta_R$ which is
expressed as $\Delta m \sim |\lambda_4+\lambda_5|\langle\phi\rangle^2/2M_{\eta^\pm}$
as found from eq.(\ref{etamass}).} 
$\eta^\pm$ become long-lived particles and their decay could be observed 
as the displaced vertex in the detector. If both $h_1$ and $|\lambda_4|$ are much smaller than 
these values, $\eta^\pm$ can be observed as heavy stable charged particles.
In the latter case,  its deviation could be larger since the Higgs-$\eta^\pm$ coupling, 
which is characterized by $\lambda_3$, can take a larger value.
These might be significant collider signatures to examine the present model.
If we combine them with the previous study of the $N_1$ DM, the promising parameter region 
could be derived. Although such a systematic study is very interesting, it is beyond the scope of the 
present work. We leave it for future work.  
  
The existence of right-handed neutrinos and the additional 
inert doublet scalar at TeV scales are expected to cause a new non-negligible contribution 
at one-loop level to LFV processes like $\mu\rightarrow e\gamma$ and also the electric dipole 
moment of electron (EDME). It is useful to examine whether their present experimental bounds 
are satisfied and also to give their prospect in the future experiments.
If we use the neutrino Yukawa couplings fixed by eqs.~(\ref{nmasseig}) and (\ref{delm}), 
we can estimate them definitely. Branching ratio of the LFV process 
$\ell_\alpha\rightarrow\ell_\beta\gamma$ in the scotogenic model is expressed as \cite{lfv} 
\begin{equation}
B(\ell_\alpha\rightarrow\ell_\beta\gamma)\simeq\frac{3\alpha}{64\pi(G_F M_\eta^2)^2}
\left|\sum_{k=1}^3U_{\alpha k}^\ast U_{\beta k}h_k^2F_2\left(\frac{M_k}{M_\eta}\right)\right|^2,
\end{equation}
and the EDME is estimated as \cite{zeromass}
\begin{equation}
d_e/e\simeq\sum_{k=1}^3\frac{U_{ek}^\ast U_{ek} h_k^2m_e}{16\pi^2M_\eta^2}
F_2\left(\frac{M_k}{M_\eta}\right)\sin\gamma_k,
\end{equation}
where $F_2(x)$ is defined as
\begin{equation}
F_2(x)=\frac{1-6x^2+3x^4+2x^6-6x^4\ln x^2}{6(1-x^2)^4}.
\end{equation}
Although these could be possible signatures of the model, 
they are respectively of $O(10^{-19})$ and $O(10^{-35})$~cm for the parameters 
given in eq.~(\ref{bench}) which are used 
to explain both the DM relic abundance and the baryon number asymmetry in the present study.
These values are much smaller than the present bounds \cite{pdg}, and they seem to be difficult to  
be reached in near future experiments. 
These results are not expected to be changed largely by varying the assumed parameters. 

\section{Summary}
The existence of both neutrino mass and DM is unexplained problem in the SM.
If it is solved in a common framework, it could be a promising candidate for new physics.
The scotogenic model is such an example. 
In this study, we propose a scenario in the original scotogenic model which 
allows the lightest right-handed neutrino to be DM without introducing extra interactions.
Since this DM cannot be detected through direct detection experiments,
such a model may be favorable from the present experimental status.  

We discuss remarkable features of the active neutrino mass matrix in the model 
and show that the right-handed neutrino with mass in a wide range from TeV to keV 
can explain the required DM abundance.
The model can also explain the baryon number asymmetry through leptogenesis based on the decay of
the right-handed neutrino if we assume mass degeneracy among $Z_2$ odd ingredients. 
Although the required mass degeneracy is strict, it is noticeable that three unsolved problems 
in the SM can be explained through the right-handed neutrinos simultaneously.
An interesting point of this scenario is that the relic abundance of DM is determined not only by 
the interaction of DM itself but also by the interaction of its mother particle $\eta_R$.
The model gives an example that the interaction searched by the direct detection 
experiments of DM is irrelevant to the interaction which determines the DM relic abundance. 

\newpage

\end{document}